\documentclass[11pt,a4paper]{article}
\usepackage{color}
\usepackage{titling}
\usepackage{amsmath, amssymb}
\usepackage{epsfig, palatino}
\usepackage{pstricks,pst-node,pst-tree}
\usepackage{epic}

\usepackage{mathrsfs}

\usepackage{ae} 
\usepackage[T1]{fontenc}
\usepackage[ansinew]{inputenc}
\usepackage{amsmath}
\usepackage{amssymb}
\usepackage{graphicx}
\usepackage{color}
\usepackage[colorlinks]{hyperref}
\usepackage{epsfig}
\usepackage{booktabs}

\newcommand{\mx}{x^{\rm mir}}
\newcommand{\px}{x^{\rm ph}}
\newcommand{\rb}{\right)}
\newcommand{\lb}{\left(}

\newcommand{\figm}{{\includegraphics[scale=0.5]{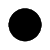}}}
\newcommand{\figf}{{\includegraphics[scale=0.5]{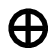}}}
\newcommand{\figF}{{\includegraphics[scale=0.5]{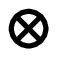}}}
\newcommand{\figb}{{\includegraphics[scale=0.5]{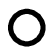}}}
\newcommand{\figp}{{\includegraphics[scale=0.5]{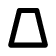}}}

\newcommand{\beq}{\begin{equation}}
\newcommand{\eeq}{\end{equation}}
\newcommand\beqa{\begin{eqnarray}}
\newcommand\eeqa{\end{eqnarray}}
\newcommand\bea{\begin{array}}
\newcommand\eea{\end{array}}

\newcommand{\im}{{\rm Im}\;}

\def\XXint#1#2#3{{\setbox0=\hbox{$#1{#2#3}{\int}$}
\vcenter{\hbox{$#2#3$}}\kern-.5\wd0}}

\newcommand{\nn}{\nonumber}

\newcommand{\COMMENT}[1]{}

\newcommand{\neqa}{\nonumber\end{eqnarray}}
\newcommand{\la}[1]{\label{#1}}

\newcommand{\eq}[1]{(\ref{#1})}

\def\tr{{\rm tr~}}

\renewcommand{\d}{\partial}

\newcommand{\<}{{\langle}}
\renewcommand{\>}{{\rangle}}

\newcommand{\re}{\relax{\rm I\kern-.18em R}}

\def\su2{{SU(2)}}

\def\eps{{\epsilon}}

\def\[{\left[}
\def\]{\right]}

\def\s{\sigma}

\def\[{\left[}
\def\]{\right]}

\def\<{\langle}
\def\>{\rangle}

\def\i2{\frac{i}{2}}

\usepackage{varioref}
\usepackage{makeidx}
\makeindex
\usepackage[english]{babel}

\begin{document}

%

\pretitle{
\begin{flushright}\footnotesize
\vspace{-2.7cm}
\texttt{DESY 09-041}\\
\texttt{LPTENS-09/09}\\
\texttt{AEI-2009-035}\\
\vspace{1.7cm}
\end{flushright}
\begin{center}\LARGE}
\posttitle{\par\end{center}\vskip 0.5em}
\preauthor{\begin{center}
\large \lineskip 0.5em%
\begin{tabular}[t]{c}}
\postauthor{\end{tabular}\par\end{center}}
\predate{\begin{center}\large}
\postdate{\par\end{center}}
\date{}

\preauthor{}
\DeclareRobustCommand{\authorthing}{
\begin{center}
\begin{tabular}{cccc}

\large {Nikolay Gromov}$^\figp$ & \large {Vladimir Kazakov}$^{\figm}$ & \large {Andrii Kozak}$^\figb$ & \large {Pedro Vieira}$^\figf$ \\
{}\\
\multicolumn{4}{c}{\footnotesize $^\figp$DESY Theory, Hamburg, Germany \& } \\
\multicolumn{4}{c}{\footnotesize II. Institut f\"{u}r Theoretische
Physik Universit\"{a}t, Hamburg, Germany \&} \\
\multicolumn{4}{c}{\footnotesize St.Petersburg INP, St.Petersburg,
Russia} \\
\\
\multicolumn{4}{c}{\footnotesize$^{\figm,\figb}$Ecole Normale Superieure, LPT,  75231 Paris CEDEX-5, France \& }\\
\multicolumn{4}{c}{\footnotesize
l'Universit\'e Paris-VI, Paris, France;} \\
\\
\multicolumn{4}{c}{\footnotesize$^\figf$Max-Planck-Institut f\"ur Gravitationphysik}\\
\multicolumn{4}{c}{\footnotesize Albert-Einstein-Institut,  14476 Potsdam, Germany  }\\
\end{tabular}
\end{center}}
\author{\authorthing}
\postauthor{}

\title{Exact Spectrum of Anomalous Dimensions of Planar $N=4$ Supersymmetric Yang-Mills Theory: TBA and excited states}
\begin{titlingpage}
\maketitle

\begin{abstract}

Using the thermodynamic Bethe ansatz  method we derive  an infinite set
of integral non-linear equations for the   spectrum of states/operators in AdS/CFT. The Y-system conjectured in \cite{Gromov:2009tv} for the spectrum of all operators in planar \(N=4\) SYM  theory follows from these equations. In particular, we present the integral TBA type equations for the spectrum of all operators within the $sl(2)$ sector.
We prove that all the kernels and free terms entering these TBA equations are real and have nice fusion properties in the relevant mirror kinematics. We find the analogue of DHM formula for the dressing kernel in the mirror kinematics.
\end{abstract}

\end{titlingpage}

\title{Spectrum of Low-Lying Exitation in AdS/CFT from TBA}

\section{Introduction}
\begin{figure}[t]
\begin{center}
\includegraphics[width=120mm]{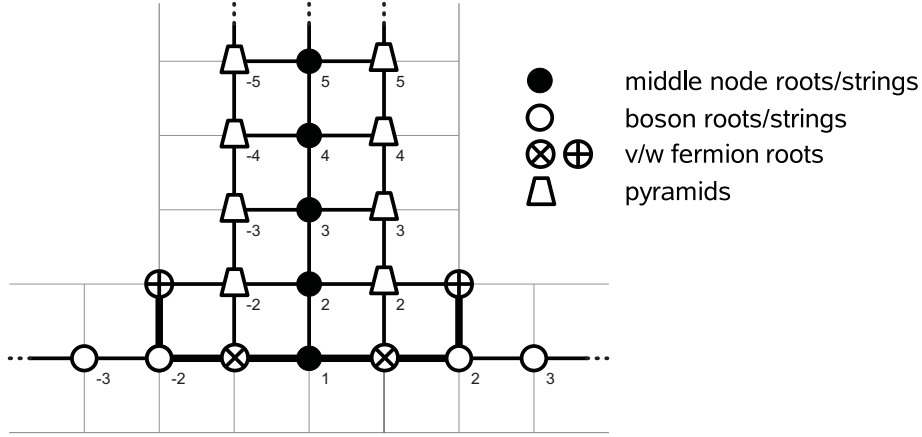}
\end{center}
\caption{\textbf{T}-shaped ``fat hook" (\textbf{T}-hook) uniting two $SU(2|2)$ fat hooks, see \cite{Kazakov:2007fy} for details on fat hooks and super algebras.
}\label{FatHook}
\end{figure}

Recently,  a set of functional equations, the so called \(Y\)-system, defining the spectrum of all local operators  in planar AdS/CFT correspondence, was proposed by three of the current authors \cite{Gromov:2009tv}. The Y-system has the form of functional equations
\begin{equation}
\label{eq:Ysystem} \frac{Y_{a,s}^+ Y_{a,s}^-}{Y_{a+1,s}Y_{a-1,s}}
 =\frac{(1+Y_{a,s+1})(1+Y_{a,s-1})}{(1+Y_{a+1,s})(1+Y_{a-1,s})} \,,
\end{equation}
where $f^{\pm}\equiv f(u\pm i/2)$ are simple shifts in the imaginary direction.
The functions \(Y_{a,s}(u)\) are defined only on the nodes marked by \(\figb,\figf,\figF,\figp,\figm\) on Fig.\ref{FatHook}. Its solutions with appropriate analytical properties define the energy of a state (anomalous dimension of an operator in N=4 SYM) through the formula\footnote{In some cases the integration contour could encircle singularities of the integrand situated away from the real axe.
In the large $L$ asymptotics these singularities can be  responsible for the L\"uscher $\mu$-terms. See also discussion in section \ref{six}.}
\begin{equation}
E=\sum_{j}\epsilon_1(u_{4,j})+\sum_{a=1}^\infty\int_{-\infty}^{\infty}\frac{du}{2\pi i}\,\,\frac{\d\epsilon_a^*}{\d u}\log\left(1+Y_{a,0}^*(u)\right) . \la{eq:Energy}
\end{equation}
where \(\eps^*_n\) is the mirror ``momentum" defined in the text below and the
rapidities $u_{4,j}$ are fixed by the exact Bethe ansatz equations
\beq
Y_{1,0}(u_{4,j})=-1\,.
\eeq

The Y-system is equivalent to the Hirota bilinear equation
\begin{equation}
 \label{Tsystem}
 T_{a,s}^+T_{a,s}^- =T_{a+1,s}T_{a-1,s}+T_{a,s+1}T_{a,s-1} \,, \\
\end{equation}
where the functions \(T_{a,s}(u)\) are non-zero only on the visible part of the 2D lattice drawn on Fig.\ref{FatHook} and
\begin{equation}
Y_{a,s}=\frac{T_{a,s+1}T_{a,s-1}}{T_{a+1,s}T_{a-1,s}}\;\; . \label{YT}
\end{equation}

It was shown that the Y-system passes a few non-trivial tests, and in particular it is completely consistent with the asymptotic Bethe  ansatz (ABA) \cite{Staudacher:2004tk,Beisert:2005fw,Beisert:2005tm,Beisert:2006ez}, is compatible with the crossing relation \cite{Janik:2006dc} and reproduces the first wrapping corrections at weak coupling for Konishi and  other twist two operators \cite{Bajnok:2008bm,Bajnok:2008qj,Fiamberti:2008sh}.

In this paper, we will provide a derivation of the $Y$-system similar in spirit to that  employed in the derivation of the TBA-type non-linear integral equations  for the finite volume spectra of relativistic 2-dimensional models. It is based on the Matsubara trick relating the ground state of a euclidean QFT on a cylinder to the free energy of the same theory in finite temperature. If we take instead of the cylinder a torus with a  small circumference \(L\)  and a large circumference \(R\) we can represent the partition function in two different channels as a sum over energiy levels. In the large \(R\) limit, we can identify the free energy \({\cal F}(L)\) per unit length of a ``mirror" QFT
living in the space section along the infinite direction of the torus and
having a temperature \(T=1/L\), with the ground state energy \(E_0(L)\) of the original QFT living on a space circle of the radius \(L\)
\begin{equation*}
Z(L,R)=\sum_{k}e^{-L\tilde E_k(R)}=\sum_{j}e^{-RE_j(L)}\to_{_{\!\!\!\!\!\!\!\!R\to\infty}} e^{-R{\cal F}(L)}= e^{-RE_0(L)}\;.
\end{equation*}

In the relativistic QFT's the original theory and the mirror theory are essentially equivalent and differ only in the boundary conditions \cite{Zamolodchikov:1991et}. An example of such a TBA calculation, useful for our further purposes, for the \(SU(2)\) principle chiral field (PCF), can be seen in the Appendix A of \cite{Gromov:2008gj}. In the supersting sigma model on \(AdS_5\times S^5\) background in the light cone gauge relevant to our problem,  we have to deal with the non-relativistic original and  mirror sigma models (see \cite{Arutyunov:2009ga,Arutyunov:2007tc}).

Particularly important for our discussion is the form of the energy and momentum of the elementary excitations for both the physical and mirror theories in infinite volume. They are conveniently parameterized in terms of the Zhukowsky variables,
\begin{equation}\label{ZHUK}
x(u)+\frac{1}{x(u)}=\frac{u}{g}
\end{equation}
which  admits two solutions, one of them outside the unit circle $|x(u)|>1$ and another inside the unit circle, $|x(u)|<1$. The energy $\epsilon_a(u)$ and momentum $p_a(u)$ of the physical bound states are then given by \cite{Dorey:2006dq}
\begin{equation}
\epsilon_a(u)= a+\frac{2ig}{x^{[+a]}}-\frac{2ig}{x^{[-a]}}\,\, , \,\, p_a(u)=\frac{1}{i}\log\frac{x^{[+a]}}{x^{[-a]}} \label{physicaldispersion}
\end{equation}
where $x^{[\pm a]}\equiv x(u\pm ia/2)$ are evaluated in the physical kinematics where $|x^{[\pm a]}| >1$.

The mirror energy and momentum are obtained by the usual Wick rotation $(E,p)\to (ip,iE)$. To stress this we denote the mirror energy by $i p_a^*$ and the mirror momentum by $i \epsilon_a^*$. The quantities \( \epsilon_a^*\) and \(p_a^*\) are defined precisely as in (\ref{physicaldispersion}) where $x^{[a]}$ are now evaluated in the mirror kinematics where $|x^{[a]}| >1$ but  $|x^{[-a]}|<1$,
for \(a>0\). 

Let us now return to our general review of the TBA method. This method is based on the so called string hypothesis: all the  eigenstates of an integrable model in the infinite volume are represented by bound states (the simplest ones are called ``strings") described by some density $\rho_A$. In terms of these densities the asymptotic Bethe equations simply read
\beq
  \bar\rho_A(u)+ \rho_A(u)=\frac{i}{2\pi}\frac{d\epsilon_A^*(u)}{du}- K_{BA}(v,u)*\rho_B(v)\,\,.
\label{eq:BAErho}
\eeq
Here $K_{BA}(v,u)=\frac{1}{2\pi i}\frac{d}{du} \log S_{AB}(u,v)$ is the kernel describing the interaction between the bound states $A$ and $B$ which scatter via an S-matrix $S_{AB}$.  $i \epsilon_A^*$ is the momentum of a magnon labeled by $A$. For the same reasons as mentioned above in the discussion of the $AdS/CFT$ dispersion relations we use  this notation to emphasize that the momenta of these mirror particles are obtained from the energy of the physical particles $\epsilon_A(u)$ by the  Wick rotation. Finally $\bar \rho_A$ is the density of holes associated  with the bound state $A$.

To compute the free energy we must minimize the functional
\beq
\mathcal{F}= \,\,\, \sum_{A}  \int_{-\infty}^\infty\,  du  \left( \left(L ip^*_A +h_A\right) \rho_{A} -  \[ \rho_A \log\left(1+\frac{\bar\rho_A}{\rho_A}\right)+  \bar\rho_A \log\left(1+\frac{\rho_A}{\bar\rho_A}\right)\right]
\right)\eeq
with respect to   \(\rho_A(u)\), \(\bar\rho_A(u) \) and  exclude \(\delta\bar\rho_A\) by the use of the constraint imposed by the BAE's \eq{eq:BAErho}. The physical origin of each term in the expression for the free energy is as follows: The first term accounts for the energy (times inverse ``temperature" $L$); the term in the square brackets represent the entropy contribution;  we  added a generic chemical potential $h_A$ for each kind of bound states. This chemical potential is needed if the theory contains fermionic excitations, as is the case for the AdS/CFT system, since we want to compute  the Witten index rather than the thermal partition function where the physical fermions are periodic. This amounts to choosing $h_A=i\pi=\log(-1)$ for the fermionic states and $h_A=0$ for the bosonic states.

The minimization of the free energy yields the TBA equations
 \beq\label{eq:TBAfirst}
\log {\cal Y}_A(u)= K_{AB}(u,v)*\log[1+1/{\cal Y}_B(v)]+iL p_A^*(u) + h_A
\eeq
for the quantities $\mathcal{Y}_A=\frac{\bar \rho_A}{\rho_A}$. Finally, at this saddle point,
the  free energy can be simply written as
\beq
\mathcal{F}= \sum_{A} \int  \frac{du}{2\pi i } \frac{d\eps_A^*}{du} \,\log\left(1+1/\mathcal{Y}_A(u) \right)   \la{fgen}  \,.
\eeq
In this way one obtains the finite volume ground state energy for a generic integrable field theory.
The excited physical states are recovered by the usual procedure of analytic continuation \cite{Dorey:1996re,Bazhanov:1996aq,Fioravanti:1996rz,Teschner:2007ng} and will be also discussed in this paper.

In what follows, we will apply the TBA method to the ``mirror" superstring sigma model and derive this AdS/CFT Y-system conjectured in  \cite{Gromov:2009tv}. The actual TBA equations arising as an intermediate  step towards the Y-system, may be very useful for the numerical calculations of the energies of low-lying states.

\section{The starting point: Beisert-Staudacher equations} \la{sec1}

The basis of our derivation of TBA for AdS/CFT are the Beisert-Staudacher (BS) ABA equations  of \cite{Staudacher:2004tk,Beisert:2005fw,Beisert:2006ez} in their mirror form \cite{Ambjorn:2005wa,Arutyunov:2007tc}. We write them in our compact notations, introducing three types of Baxter functions
\beq
\la{RBQ}  R_l^{(\pm)}(u)\equiv \prod_{j=1}^{K_l} \frac{x(u)-x_{l,j}^{\mp}}{\sqrt{x^{\mp}_{l,j}}},\quad  B_l^{(\pm)}(u)\equiv\prod_{j=1}^{K_l} \frac{\frac{1}{x(u)}-x_{l,j}^{\mp}}{\sqrt{x^{\mp}_{l,j}}}  ,\quad Q_l(u)=\prod_{j=1}^{K_l}(u-u_{l,j})=(-g)^{K_l}R_{l}(u)B_{l}(u)  .
\eeq
The index $l$ takes the values $l=1L,2L,3L$ or $l=1R,2R,3R$ parametrizing the rapidities of the left and right $SU(2|2)$ wings of the model, correspondingly.
\(R^{(\pm)}\) and \(B^{(\pm)}\) with no subscript $l$ correspond to the roots \(x_{4,j}\) of the middle node and \(R_{l},B_l\) without supercript $(+)$ or $(-)$ are defined as in (\ref{RBQ}) with $x^\pm_j$ replaced by $x_j$. In these notations the left wings ABA's read:
\begin{eqnarray}
\left. 1= \frac{Q^+_{2L}B^{(-)}}{Q_{2L}^-B^{(+)}}\right|_{u_{1L,k}}\,,\qquad\!  \label{BAE}
\left. -1=\frac{Q_{2L}^{--}Q_{1L}^+Q_{3L}^+}{Q_{2L}^{++}Q_{1L}^-Q_{3L}^-}\right|_{u_{2L,k}}\,,\qquad
\left. 1= \frac{Q_{2L}^+ R^{(-)}}{Q_{2L}^- R^{(+)}}\right|_{u_{3L,k}}
\end{eqnarray}
with a similar set of equations for the right wing replacing $L\to R$. The Bethe equation for the  middle node for the full AdS/CFT ABA of \cite{Beisert:2005fw} fix the positions of the $u_{4,j}$ roots
 from\footnote{This equation is identical to the eq.(6.6) from \cite{Arutyunov:2007tc}.
 The factors of $x^+/x^-$ outside of the square brackets can be easily reconstructed from the
 unimodularity of the r.h.s. of \eq{middle}. We thank the referee for pointing us out this misprint
 which fortunately does not affect any of our results in the previous version of the preprint.
}
\beq
\label{middle}
-1\!=\left.\left[e^{R\, \epsilon_1^*}
\left(\!\frac{Q_4^{--}}{Q_4^{++}}\frac{B_{1L}^{+}  R_{3L}^{+}}{B_{1L}^{-}
 R_{3L}^{-}}\frac{B_{1R}^{+}  R_{3R}^{+}}{B_{1R}^{-}
 R_{3R}^{-}}\!\right)^{\!\!}
\left(\!\frac{ B^{+(+)}}{B^{-(-)}}\!\right)^{\!\!2}  S^2\, \right]
\prod_{j=1}^{K_4}\frac{x_{4,j}^+}{x_{4,j}^-}
\left(\frac{x_{4,k}^+}{x_{4,k}^-}\right)^{\frac{K_{1R}-K_{3R}+K_{1L}-K_{3L}}{2}}\right|_{u=u_{4,k}}
\eeq
for the \(sl(2) \) favored grading. The dressing factor is
$S(u)=\prod_j\sigma(x(u),x_{4,j})$ where $\sigma$ is the BES dressing kernel \cite{Beisert:2006ez}  (see \cite{Dorey:2007xn} for a nice integral representation of the dressing kernel).

\section{Bound states and TBA equations for the mirror ``free energy"} \la{sec2}
 To write the TBA for the full AdS/CFT, we have to find the BAE's for
the densities of all complexes of Bethe roots in the infinite volume
$R=\infty$. The string hypothesis implies the full description  of the infinite volume solutions. They are easy to classify: there is only one type of
momentum carrying complexes, strings in the middle nodes, similar to
standard $SU(2)$ strings \cite{Dorey:2006dq}; the rest are the same
complexes as found by Takahashi in the Hubbard model
\cite{Takahashi,KorepinBook} (see also \cite{Arutyunov:2009zu}).

As the result, we find that in the large \(R\)  limit of BAE's the roots regroup into the following bound states:
\begin{eqnarray*}\label{Table1}
&& u_{4}\,\, \,\,\,\,=u+ij\,\,  , \,\,  j=-\frac{n-1}{2},\dots,\frac{n-1}{2} \,\, :\quad \,\, \text{middle node bound states}\qquad : \figm_n   \\ \hline \\
&&\vspace{1mm} u_{2}^{L,R}=u+ij \,\,, \,\,  j=-\frac{n-2}{2},\dots,\frac{n-2}{2} \,\, :\quad \,\, L,R  \text{ string bound states} \,\,\,\,\qquad: \figb_{\pm n} \\ \hline \\
&&u_{3}^{L,R}=u+ij \,\,, \,\,  j=-\frac{n-1}{2},\dots,\frac{n-1}{2}\,\,\quad \\
&&u_{2}^{L,R}=u+ij \,\,, \,\,  j=-\frac{n-2}{2},\dots,\frac{n-2}{2} \,\, :\quad \,\ L,R \text{ trapezia} \qquad\qquad\qquad\quad\,\,:\,\,\figp_{\pm n}
\\
&&u_{1}^{L,R}=u+ij \,\,, \,\,  j=-\frac{n-3}{2},\dots,\frac{n-3}{2}
  \\ \hline \\
&&u_{1}^{L,R}=u  \,\, \,\,\,\,\,\,\,\,\,\,\, \,\,\,\,\,\,\,\, \,\,\,\, \,\,\,\,\,\,\,\,\, \,\,\,\,\,\,\,\,\, \,\, \,\,\,\,\,\,\,\,\, \,\,\,\,\,\,\,\,\, \,\,\,\,\,\,\,\,\, \,\,\,\,\,\,\,\,\, : \,\, \quad L,R \text{ single fermion} \,\,\,\,\,\,\,\,\,\,\,\,\,\, \,\,\,\,\,\,\,\,\, \,\,\,\,\,\,:\,\,{\figf_{\pm}}
  \\ \hline \\
&&u_{3}^{L,R}=u  \,\, \,\,\,\,\,\,\,\,\,\,\, \,\,\,\,\,\,\,\, \,\,\,\, \,\,\,\,\,\,\,\,\, \,\,\,\,\,\,\,\,\, \,\, \,\,\,\,\,\,\,\,\, \,\,\,\,\,\,\,\,\, \,\,\,\,\,\,\,\,\, \,\,\,\,\,\,\,\,\, : \,\, \quad L,R \text{ single fermion} \,\,\,\,\,\,\,\, \,\,\,\,\,\,\,\,\,\,\,\,\,\,\, \,\,\,\,\,\,: \,\, {\figF_\pm}
\end{eqnarray*}
where by \(u\) we  denote the real center of a complex. Thus the index $A$ in formulae (\ref{eq:BAErho}-\ref{fgen}) takes the values
\beq
A=\{ \figb_{\pm n},\figf_{\pm},\figF_{\pm},\figp_{\pm n},\figm_n\}
\eeq
or, in the notation used in \cite{Gromov:2009tv} for the points on the T-hook\beq
A=\{ (1,\pm n),(2,\pm 2),(1,\pm 1),(n,\pm 1),(n,0)\}\;.
\eeq
Multiplying the Bethe equations along each complex we obtain the fused equations (\ref{eq:BAErho}) for the densities (of particles and holes,  \(\rho_A(u)\) and \(\bar\rho_A(u) \)) of the centers of complexes (\ref{eq:TBAfirst}). It is useful to introduce the following notation for $\mathcal{Y}_A$:
\begin{eqnarray}
\left\{\mathcal{Y}_{\figb_{\pm n}},\mathcal{Y}_{\figf_{\pm}},\mathcal{Y}_{\figF_{\pm}},\mathcal{Y}_{\figp_{\pm n}},\mathcal{Y}_{\figm_{\pm n}}\right\}
&=&\left\{{Y}_{\figb_{\pm n}} ,Y_{\figf_\pm},\frac{1}{Y_{\figF_\pm}},\frac{1}{Y_{\figp_{\pm n}}},\frac{1}{Y_{\figm_{\pm n}}  }\right\}
\la{notation}
\end{eqnarray}
In particular notice that the $Y$ functions $Y_{a,s}$ arrange nicely into a T-shaped form as depicted in Fig.\ref{FatHook}. As shown below, these functions are precisely those appearing in the $Y$-system (\ref{eq:Ysystem}).

The only complexes which carry energy and momentum are those made out of middle node roots $u_{4,j}$,
\beq
 \eps_A^*=\delta_{A,{\figm_n}} \epsilon_n^* \,\, , \,\,  p_A^*=\delta_{A,{\figm_n}} p_n^*
\eeq
where $\epsilon_{n}^*$ and $p_n^*$ are explained after (\ref{physicaldispersion}). The fused kernels $K_{AB}$ are given by
 \beqa\la{Kern}
K_{AB}=
\bea{|l||l|l|l|l|l|}\hline
A\backslash B &\figb_m&\figf_+&\figF_+&\figp_m&\figm_m\\ \hline\hline
\figb_n & +K_{n-1,m-1} &-K_{n-1}&+K_{n-1}&0&0\\ \hline
\figf_+& -K_{m-1} & 0 & 0&+K_{m-1}& -\mathcal{B}^{(01)}_{1m} \\ \hline
\figF_+& -K_{m-1} & 0 & 0 &+K_{m-1}&- \mathcal{R}^{(01)}_{1m} \\ \hline
\figp_n &0& -K_{n-1}&+K_{n-1}&+K_{n-1,m-1}&-{\cal R}_{nm}^{(01)}-{\cal B}_{n-2,m}^{(01)}
\\ \hline
\figm_n
&0&\mathcal{B}^{(10)}_{n1}&-\mathcal{R}^{(10)}_{n1}&-\mathcal{R}^{(10)}_{nm}-\mathcal{B}^{(10)}_{n,m-2}
& -2\mathcal{S}_{nm}-\mathcal{B}_{nm}^{(11)}+\mathcal{R}_{nm}^{(11)}\\ \hline
\eea
\eeqa
where the block entrees of this infinite matrix are defined as
\beqa
&&K_{n}\equiv \frac{1}{2\pi i}\frac{d}{dv} \log
\frac{ u-v+in/2}{ u-v-in/2}  \, , \quad
K_{nm}\equiv
\sum_{j=-\frac{m-1}{2}}^{\frac{m-1}{2}}\sum_{k=-\frac{n-1}{2}}^{\frac{n-1}{2}}
K_{2j+2k+2}\\
&&{\cal S}_{nm}(u,v)\equiv  \frac{1}{2\pi i}\frac{d}{dv} \log \sigma(x^{\pm n}(u),x^{\pm
m}(v)) \\
&&
\mathcal{B}^{(ab)}_{nm}(u,v)\equiv
\sum_{j=-\frac{n-1}{2}}^{\frac{n-1}{2}}\sum_{k=-\frac{m-1}{2}}^{\frac{m-1}{2}}
\frac{1}{2\pi i}\frac{d}{dv} \log
\frac{b(u+ia/2+ij,v-ib/2+ik)}{b(u-ia/2+ij,v+ib/2+ik)}\\
&&\mathcal{R}^{(ab)}_{nm}(u,v)\equiv
\sum_{j=-\frac{n-1}{2}}^{\frac{n-1}{2}}\sum_{k=-\frac{m-1}{2}}^{\frac{m-1}{2}}
\frac{1}{2\pi i}\frac{d}{dv} \log
\frac{r(u+ia/2+ij,v-ib/2+ik)}{r(u-ia/2+ij,v+ib/2+ik)}
\eeqa
where
\beq
r(u,v)=\frac{x(u)-x(v)}{\sqrt{x(v)}}\;\;,\;\;b(u,v)=\frac{1/x(u)-x(v)}{\sqrt{x(v)}}\;. \la{brs}
\eeq
In the table above we only wrote the interaction between the complexes of the left $SU(2|2)$ wing, between those complexes and the middle node bound states, as well as between the middle node bound states themselves. The right wing interaction is of course absolutely identical and the complexes of different wings do not interact. Equations (\ref{eq:TBAfirst}) in the notation of  (\ref{notation}) then read
\beqa\la{eq:YF}
\log Y_{\figF_\pm }&=&
+K_{m-1}*\log\frac{1+1/Y_{\figb_{\pm m}}}{1+Y_{\figp_{\pm m}}}
+\mathcal{R}_{1m}^{(01)}*\log(1+Y_{\figm_m})
+\log(-1)
\\ \la{eq:Yf}
\log Y_{\figf_\pm }&=&
-K_{m-1}*\log\frac{1+1/Y_{\figb_{\pm m}}}{1+Y_{\figp_{\pm m}}}
-\mathcal{B}_{1m}^{(01)}*\log(1+Y_{\figm_m})-
\log(-1)
\\ \la{eq:Yp}
\log Y_{{\figp}_{\pm n}}&=&-K_{n-1,m-1}*\log(1+Y_{{\figp}_{\pm m}})
-K_{n-1}*\log\frac{1+Y_{\figF_\pm }}{1+1/Y_{\figf_\pm }} \\ &+&\left({\cal R}^{(01)}_{nm}+{\cal B}^{(01)}_{n-2,m}\right)*\log(1+Y_{\figm_m})\nn
\\ \la{eq:Yb}
\log Y_{{\figb}_{\pm n}}&=&K_{n-1,m-1}*\log(1+1/Y_{{\figb}_{\pm m}})
+K_{n-1}*\log\frac{1+Y_{\figF_\pm}}{1+1/Y_{\figf_\pm}} \\ \la{eq:Ym}
\log Y_{{\figm}_{n}}&=&
L\log\frac{x^{[-n]}}{x^{[+n]}}+\left(2{\cal S}_{nm}-{\cal R}_{nm}^{(11)}+{\cal B}_{nm}^{(11)}\right)*\log(1+Y_{\figm_m})\\
&-&{\cal B}_{n1}^{(10)}*\log(1+1/Y_{\figf_+})
+{\cal R}_{n1}^{(10)}*\log(1+Y_{\figF_+})
+\left({\cal R}_{nm}^{(10)}+{\cal B}_{n,m-2}^{(10)}\right)*\log(1+Y_{\figp_m})\nn\\
&-&{\cal B}_{n1}^{(10)}*\log(1+1/Y_{\figf_-})
+{\cal R}_{n1}^{(10)}*\log(1+Y_{\figF_-})
+\left({\cal R}_{nm}^{(10)}+{\cal B}_{n,m-2}^{(10)}\right)*\log(1+Y_{\figp_{-m}})\nn
\eeqa
All convolutions are to be understood in the usual sense with the second variable being integrated over so that $K*f=\int dv K(u,v)  f(v)$. Summation over the repeated index $m$ is assumed  ($m=2,\dots,\infty$ for the convolutions involving  pyramids $\figp_{\pm m}$ and strings $\figb_{\pm m}$ and  $m=1,\dots,\infty$ for the convolutions with the middle node bound states $\figm_m$). There are still some ambiguities involved in these integral equations concerning the choice of the integration contours. We will discuss this, still not completely elucidated, point when we will consider  equations for the excited states where some of the ambiguities will be lifted.

\section{Derivation of the AdS/CFT Y-system}

We will now derive, from the TBA equations, the Y-system \eq{eq:Ysystem} and  \eq{Tsystem} for the AdS/CFT spectrum conjectured in  \cite{Gromov:2009tv}. We shall do it separately for each type of excitations.

The key idea in the derivation is to use the discrete Laplace operator acting on the free variable $u$ and free index $n$ in the TBA equations. We notice that
\beq
\Delta K_n(u)\equiv K_n(u+i/2-i0)+K_n(u-i/2+i0)-K_{n+1}(u)-K_{n-1}(u)=\delta_{n,1}\delta(u) \nn
\eeq
As a simple  consequence of this identity we find
\beqa\nn
\Delta K_{nm}(v-u)&=&\Delta{\cal R}^{(11)}_{nm}(v,u)=\delta_{n,m+1}\delta(v-u)+\delta_{n,m-1}
\delta(v-u)\\ 
\Delta{\cal R}^{(01)}_{nm}(v,u)&=&\Delta{\cal R}^{(10)}_{nm}(v,u)=\delta_{n,m}\delta(v-u) \label{eq:LAP1}
\eeqa
whereas the Laplacian kills all other kernels, $\Delta \mathcal{S}_{nm}=0$, etc. For example, the fact that the dressing factor is killed by the Laplacian follows from its harmonic form
\beq
\sigma_{nm}(u,v)=e^{\chi(u+in/2,v+in/2)+\chi(u-in/2,v-in/2)-\chi(u-in/2,v+in/2)-\chi(u+in/2,v-in/2)} \la{Slap}
\eeq without any singularities in the physical kinematics (this fact was already used in \cite{Gromov:2009tv} when constructing the large $L$ solutions of the $Y$-system).
By virtue of these identities we can easily compute the combinations $\log\frac{Y_{{\figb}_{n}}^+Y_{{\figb}_{n}}^-}{Y_{{\figb}_{n+1}}Y_{{\figb}_{n-1}}}$, $\log\frac{Y_{{\figp}_{n}}^+Y_{{\figp}_{n}}^-}{Y_{{\figp}_{n+1}}Y_{{\figp}_{n-1}}}$ and $\log\frac{Y_{{\figm}_{n}}^+Y_{{\figm}_{n}}^-}{Y_{{\figm}_{n+1}}Y_{{\figm}_{n-1}}}$, where $f^{\pm}\equiv f(u\pm i/2 \mp i0)$, using respectively (\ref{eq:Yb}), (\ref{eq:Yp}) and (\ref{eq:Ym}). We find
\beq
\log\frac{Y_{{\figb}_{n}}^+Y_{{\figb}_{n}}^-}{Y_{{\figb}_{n+1}}Y_{{\figb}_{n-1}}}
=\log(1+1/Y_{{\figb}_{n+1}})(1+1/Y_{{\figb}_{n-1}}) \,\, , \,\, n>2
\eeq
and
\beq
\log\frac{Y_{{\figb}_{2}}^+Y_{{\figb}_{2}}^-}{Y_{{\figb}_{3}}}
=\log\frac{(1+Y_{\figF_+})(1+1/Y_{{\figb}_{3}})}{1+1/Y_{\figf_+}}\\
\eeq
for the string bound states. The equations for  \(Y_{1,n}\) at \(n\le -2,\)  as well as their derivation, are similar. For the pyramid complexes we obtain
\beq
\log\frac{Y_{{\figp}_{n}}^+Y_{{\figp}_{n}}^-}{Y_{{\figp}_{n+1}}Y_{{\figp}_{n-1}}}
=\log\frac{1+Y_{\figm_n}}{(1+Y_{{\figp}_{n+1}})(1+Y_{{\figp}_{n-1}})} \,\, , \,\, n>2
\eeq
and
\beqa\label{PYRPYR}
\log \frac{Y_{\figp_2}^+ Y_{\figp_2}^-}{ Y_{\figp_3}}
=\log\frac{(1+Y_{\figf})(1+Y_{\figm_2})Y_{\figF_+}}{(1+Y_{\figp_3})(1+Y_{\figF_+})}-\log
Y_{\figF_+}Y_{\figf_+}+\sum_n (\mathcal{R}^{(01)}_{n1}-\mathcal{B}^{(01)}_{n1})*\log(1+Y_{\figm_n})\nn\;.
\eeqa
The first term in the r.h.s. of this equation reproduces again the correct structure of the Y-system (\ref{eq:Ysystem}). In fact, we will see below that the last two terms cancel each other and hence this equation  perfectly fits the Y-system   \eq{eq:Ysystem}. Finally, for the middle node bound states, we kill again the  kernels when applying the discrete Laplace operator and  obtain
\beq
\log \frac{Y_{{\figm}_{n}}^+Y_{{\figm}_{n}}^-}{Y_{{\figm}_{n+1}}Y_{{\figm}_{n-1}}}
=\log\frac{(1+Y_{\figp_n})(1+Y_{\figp_{-n}})}{(1+Y_{\figm_{n+1}})(1+Y_{\figm_{n-1}})} \,\, , \,\, n>1
\eeq
and
\beq
\log \frac{Y_{{\figm}_{1}}^+Y_{{\figm}_{1}}^-}{Y_{{\figm}_{2}}}
=\log\frac{1+Y_{\figF_+}}{1+Y_{\figm_{2}}}  \,.
\eeq
We are left with the equations for the two fermionic nodes \(Y_{1,1}= Y_{\figF_+}\) and \(Y_{2,2}=Y_{\figf_+}\)  (for \(Y_{1,-1}\) and \(Y_{2,-2}\) it will be similar). We consider first the node $Y_{1,1}$. Combining equations (\ref{eq:YF}) for $u\to u\pm i/2 \mp i0$ with equations (\ref{eq:Yp}) and (\ref{eq:Yb}) for real $u$ and $n=2$ we obtain (again using the fusion properties of several kernels),
\beq
\log\frac{Y_{\figF_+}^+
Y_{\figF_+}^-}{Y_{\figp_2}Y_{\figb_2}}=\log\frac{(1+1/Y_{\figb_2})(1+Y_{\figm_1})}{1+Y_{\figp_2}}
\eeq
perfectly reproducing the the equation for  \(Y_{1,1}\) from the \(Y\)-system \eq{eq:Ysystem}. Finally, to find the equation for the last fermion node $Y_{2,2}$ we simply add up equations (\ref{eq:Yf}) and (\ref{eq:YF}) to get
\beq\label{eq:Y22Y11}
\log Y_{\figF_+}Y_{\figf_+}=\sum_m \left(\mathcal{R}_{1m}^{(01)}-\mathcal{B}_{1m}^{(01)}\right)*\log(1+Y_{\figm_m})
\eeq
This shows indeed that the two last terms in \eqref{PYRPYR} cancel.
The equation for \(Y_{22}=Y_{\figf_+}\) is not a part of \(Y\)-system \eq{eq:Ysystem}  since in the standard form  it would contain the ratio $\frac{1+Y_{23}}{1+1/Y_{32}}=\frac{0}{0}$. It is thus natural that one can not render this equation local if we only use the finite $Y$ functions, see also \cite{Juttner:1997tc}. However, in terms of the T-functions appearing in \ref{YT} we believe, and partially checked, that Hirota equation \ref{Tsystem} is well defined on the full T-shaped fat-hook of figure \ref{FatHook}.

All these  equations precisely reproduce the $Y$-system (\ref{eq:Ysystem}) under the identification
\begin{eqnarray}
\left\{{Y}_{\figb_{\pm n}} ,Y_{\figf_\pm},{Y_{\figF_\pm}},{Y_{\figp_{\pm n}}},{Y_{\figm_{\pm n}}  }\right\}
=\left\{Y_{1,\pm n},{Y}_{2,\pm 2},{{Y}_{1,\pm1}},{{Y}_{n,\pm 1}},{{Y}_{n,0}}\right\}
\end{eqnarray}
mentioned in the previous section!

\section{ Integral equations for excited states}
In this section we will consider the non-linear integral TBA-type equations for  excited states.
For simplicity we shall consider only  the states in the $SL(2)$ sector, corresponding
to operators of the form $\tr (D^S Z^J )+\rm permutations$. Notice that since none of
the wings are excited the $Y$-functions will have the symmetry $Y_{a,s}=Y_{a,-s}$ which
also  means  that $Y_{\figF_+}=Y_{\figF_-}\equiv Y_{\figF},\dots$. To consider such excited
states we employ the standard analytic continuation trick
\cite{Dorey:1996re,Bazhanov:1996aq,Teschner:2007ng} where we pick extra singularities
in the convolutions with $Y_{\figm_1}$ at the points where $Y_{\figm_1}(u_{4,j})=-1$.
This procedure contains some ambiguities and the result should
be considered as a conjecture.
In this way, the free energy (\ref{fgen}) becomes (\ref{eq:Energy}) while the non-linear
integral equations of section (\ref{sec2}) are modified by the terms in the square brackets
\beqa\label{TBAeqs}
\log Y_{\figF}&=&
+K_{m-1}*\log\frac{1+1/Y_{\figb_{m}}}{1+Y_{\figp_{m}}}
+\mathcal{R}_{1m}^{(01)}*\log(1+Y_{\figm_m})
+ \left[\log\frac{R^{(+)}}{R^{(-)}} \la{Yeq1}\right]
+\log(-1)
\\
\log Y_{\figf}&=&
-K_{m-1}*\log\frac{1+1/Y_{\figb_{m}}}{1+Y_{\figp_{ m}}}
-\mathcal{B}_{1m}^{(01)}*\log(1+Y_{\figm_m})
-
\left[\log\frac{B^{(+)}}{B^{(-)}}\right]
-\log(-1)
\\
\log Y_{{\figp}_{ n}}&=&-K_{n-1,m-1}*\log(1+Y_{{\figp}_{ m}})
-K_{n-1}*\log\frac{1+Y_{\figF}}{1+1/Y_{\figf}}\nn+\left({\cal R}^{(01)}_{nm}+{\cal B}^{(01)}_{n-2,m}\right)*\log(1+Y_{\figm_m})\\
&+&\left[\sum_{k=-\frac{n-1}{2}}^{\frac{n-1}{2}}\log\frac{R^{(+)}(u+i k)}{R^{(-)}(u+i
k)}+\sum_{k=-\frac{n-3}{2}}^{\frac{n-3}{2}}\log\frac{B^{(+)}(u+i k)}{B^{(-)}(u+i
k)}\right] \la{YEQp}
\\ \la{eqlast}
\log Y_{{\figb}_{ n}}&=&K_{n-1,m-1}*\log(1+1/Y_{{\figb}_{ m}})
+K_{n-1}*\log\frac{1+Y_{\figF}}{1+1/Y_{\figf}}\\
\la{Yeq2} \log Y_{{\figm}_{n}}&=&
L\log\frac{x^{[-n]}}{x^{[+n]}}+\left(2{\cal S}_{nm}-{\cal R}_{nm}^{(11)}+{\cal B}_{nm}^{(11)}\right)*\log(1+Y_{\figm_m})+
\left[\sum_{k=-\frac{n-1}{2}}^{\frac{n-1}{2}}i\Phi(u+ik)\right]\\
&+&2\left(
{\cal R}_{n1}^{(10)}*\log(1+Y_{\figF})- {\cal B}_{n1}^{(10)}*\log(1+1/Y_{\figf})
+\left({\cal R}_{nm}^{(10)}+{\cal B}_{n,m-2}^{(10)}\right)*\log(1+Y_{\figp_{ m}})\right)\nn
\eeqa
where
\beq\label{Phy function}
\Phi(u)=\frac{1}{i}\log\left(S^2\frac{B^{(+)+}R^{(-)-}}{B^{(-)-}R^{(+)+}}\right)\;.
\eeq
and $B$ and $R$ and $S$ containing the positions of  rapidities of the excited states are defined in section \ref{sec1}. These rapidities are constrained by the \textit{exact} Bethe equations
\beq
Y_{\figm_1}(u_{4,j})=-1\,\, , \,\, j=1,\dots,M\,.
\eeq
In the convolutions involving the fermionic $Y$-functions $Y_{\figf}$ and $Y_{\figF}$ we integrate over $v \in [-2g,2g]$
\footnote{Another possibility, consistent with the infinite length solution of \cite{Gromov:2009tv}, is to choose $v \in ]-\infty,-2g]\cup[2g,\infty[ $. We will examine
that possibility in detail in the next section. We thank G.~Arutyunov and
S.~Frolov for the correspondence on this issue.  }.
We found that prescription to be consistent with the asymptotical large $L$ solution of the
Y-system derived in \cite{Gromov:2009tv}.
 In fact as one can see from these integral equations we can think of the two functions $Y_\figf$ and $1/Y_{\figF}$ as two branches of the same function. In this language the convolutions can be recasted into some nice $B$-cycle contour integrals in the $x(u)$ Riemann sheet. This is reminiscent of the inversion symmetry in the BS equations which allows one to reduce the seven Bethe equations to a smaller set of five equations \cite{Beisert:2005fw}.

An important check of these equations is the limit where $L\to\infty$. The solution of the Y-system in this limit was constructed explicitly in \cite{Gromov:2009tv}. We checked numerically that for large $L$ our integral equations are consistent with the large volume solution.

\section{Physical and mirror  choices of branches}\label{reality}
The above system of TBA equations should be valid for any value of the spectral
parameter \(u\) and it should be possible to analytically continue it to any point of the Riemann surface of the multi-valued \(Y\)-functions. But
the choice of branches to formulate the TBA equations can be very important
for its good definition and in particular for the future numerical applications. In this section we will fix a particular choice of
branches in the kernels involved in the integral equations.
This  choice will be quite unique, with the following nice properties for the \(Y\)-functions and the integration kernels:
\begin{itemize}
\item \ They  have only a minimal number of cuts,  in general only a
pair of cuts, which means that they obey an ordinary fusion procedure where all the intermediate constituents of a bound state but the first and the
last  cancel.

\item\ They are real functions of the spectral parameter \(u\) on the real axis. It fits well their physical meaning
in TBA as of the ratios of densities of physical particles and holes.\(\)

\end{itemize}

These properties will stem of course from the similar properties of integration kernels
and free terms (with no convolutions) in the TBA equations \eqref{TBAeqs}-\eqref{Yeq2}.

There are two natural possibilities to define $x(u)$ compatible with (\ref{ZHUK}).
We define two functions
\beq\label{two branches}
\px(u)=\frac{1}{2}\lb \frac{u}{g}+\sqrt{\frac{u}{g}-2}\;\sqrt{\frac{u}{g}+2}\rb\;\;,\;\;
 \mx(u)=\frac{1}{2}\lb \frac{u}{g}+i\sqrt{4-\frac{u^2}{g^2}}\rb \,.
\eeq
They both solve \eq{ZHUK}. It is easy to check that with this choice of branches  \eqref{physicaldispersion}
reproduces the physical and mirror dispersion relations,
correspondingly \cite{Arutyunov:2007tc}.
They coincide  above the real axes and have the following properties under complex
conjugation
\beq
\overline \px=\px\;\;,\;\;\overline  \mx=1/\mx\;.
\eeq
Basically both  representations \eqref{two branches} describe  the same function,
with the same Riemann surface but extended from the upper half plane
to the plane with the cut $(-2g,2g)$ for \(x^{\rm ph}\), and to the plane with
the infinite cut
$(-\infty,-2g)\cup (2g,\infty)$ for the function $\mx$.
One can say that they are two sections of the same Riemann surface.

We can plot them in Mathematica by running e.g. \\
{\small
\verb"z=a+b I;xmr=1/2(z+I Sqrt[4-z^2]);xph=1/2(z+Sqrt[z-2]Sqrt[z+2]);"\\
\verb"Plot3D[{Im[xph],Im[xmr]+0.1},{a,-3,3},{b,-1,1},PlotStyle->{Red,Yellow}]"  }

Notice that in the mirror ABA \cite{Arutyunov:2007tc} \eqref{BAE} and \eqref{middle} which we started from, the choice $\mx$ is employed
\cite{Ambjorn:2005wa,Arutyunov:2007tc}. However, for the physical ABA of Beisert and Staudacher \cite{Beisert:2005fw} we only use the physical choice $\px$. Thus to have a link with the ABA in the physical channel one should use the same definition \eq{RBQ} with
\beq
x_{j}\equiv \px(u_j)\;\;,\;\;x_{j}^{\pm}\equiv \px(u_j\pm i/2)\;,
\eeq
in  various free terms (with no convolutions) in the TBA equations.

On the other hand, since all the kernels in the TBA equations are coming from the mirror theory, both arguments should be in  mirror kinematics. Hence we specify in definitions \eq{brs}
for the integration kernels the following branches\footnote{the same branches are used in \cite{Bombardelli:2009ns,Arutyunov:2009ur}}
\beq
r(u,v)=\frac{\mx(u)-\mx(v)}{\sqrt{\mx(v)}}\;\;,\;\;b(u,v)=\frac{1/\mx(u)-\mx(v)}{\sqrt{\mx(v)}}\;.
\eeq
With this choice of branches, it is easy to check that the kernels ${\cal R}_{nm}^{(ab)},\;{\cal B}_{nm}^{(ab)}$ entering
our TBA integral equations \eq{TBAeqs}-\eq{Yeq2} are all real!
In the next section we show that  the kernel involving the dressing factor,
\(2{\cal S}_{nm}\),
is real as a consequence of crossing, up to a simple square root factor which we identify there. Moreover, together with ${\cal
R}_{nm}^{11}-{\cal B}_{nm}^{11}$ appearing in (\ref{Yeq2}), it has very simple analytic properties. Namely, it has only four branch points for each of two variables, confirming the nice fusion property announced above. We will also present a simple integral representation for this combination.
\begin{figure}
\includegraphics[scale=1]{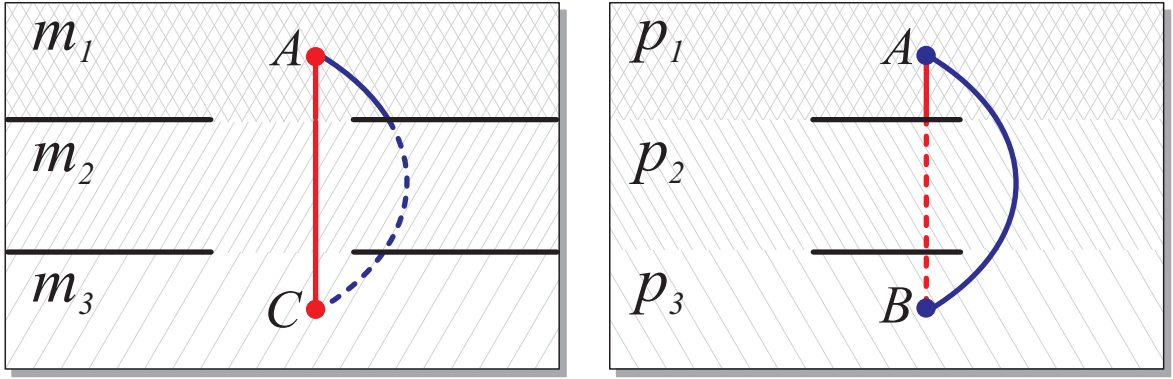}
\caption{\it Structure of the cuts and conjugation paths on the mirror and physical sheets.}\label{Mir-Phys}
\end{figure}

\subsection{Reality and crossing}
 One of the important consequenses of the crossing for the
\(SU(2)\times SU(2)\) principal chiral  model considered
in \cite{Gromov:2008gj} was the reality of the function $Y_0$
corresponding  to the single middle node in  the finite size TBA
equations. Here we show that exactly the same fenomenon is taking
place  in the present  AdS/CFT TBA equations.

Similarly to the Beisert-Eden-Staudacher physical dressing factor, the
mirror S-matrix ought to be a pure phase. Let us here explain why this
follows indeed in a simple way from the crossing relation for the
dressing factor. The same argument can be easily adapted to prove that
the leading large volume $Y$-functions found in \cite{Gromov:2009tv}
are indeed real.

We present schematically the mirror and physical sheets on the
figures \ref{Mir-Phys}.
They are naturally devided by  cuts into three regions denoted by
\(m_1,m_2,m_3\)
and   \(p_1,p_2,p_3\), correspondingly.
Since $x^{ph}(u)$ coincides with $x^{mr}(u)$ in the upper half-plane
the regions $p_1$ and $m_1$ are equivalent,  \(p_1=m_1\).

Let us consider two points $u_A$ and $v_A$ above the upper cut, i.e in
the region \(p_1=m_1\).
Conjugation in the mirror sheet sends these points to $u_B\equiv \tilde
u_A$ and $v_B  \equiv\tilde v_A$ (belonging to the physical sheet)
while conjugation in the physical sheet maps them to $u_C\equiv \bar
u_A$ and $v_C\equiv \bar v_A$ (belonging to the mirror sheet).

Notice that crossing condition relates the dressing factor with
argument $u_B$ with the dressing factor at the point $u_C$. More
precisely, we have \cite{Janik:2006dc}
\beq
\sigma(u_B,v_B)\sigma(u_C,v_B)=\frac{y^-}{y^+}
\frac{x^--y^-}{x^+-y^-}
\frac{1/x^--y^+}{1/x^+-y^+}\;\;,\;\;x=\px(u_B)\;\;,\;\;y=\px(v_B)
\eeq
Notice also that we can now analytically continue both sides of this
equality with respect to the $v_B$ root, in particular we can generate
the crossing relation where $v_B$ is replaced by $v_C$.
Using again the (analytically continued) crossing relation to
transform $v_B$ into $v_C$   we get
\beq\la{cr2}
\sigma(u_C,v_C)
=\frac{x^- y^+}{x^+ y^-}\sigma(u_B,v_B) \;\;,\;\;x=\px(u_B)\;\;,\;\;y=\px(v_B)
\eeq
Taking the complex conjugate of this expression and using the fact that
the dressing factor is a pure phase on the physical sheet we get
\cite{Arutyunov:2007tc}
\beq\la{cr2}
\left(\sigma(\bar u_A,\bar v_A)\right)^*
=\frac{x^+ y^-}{x^- y^+} \frac{1}{\sigma(u_A,v_A)}
\;\;,\;\;x=\mx(u_A)\;\;,\;\;y=\mx(v_A)
\eeq
Notice that we replaced $\px(u_A)$ and $\px(v_A)$ by their mirror
counterparts because $A$ is in the region $p_1=m_1$. Furthermore, in
the left hand side, we explicitly wrote \(u_C=\bar u_A\) and
\(v_C=\bar v_A\) to recognize  the explicit definition of the
conjugated function on the mirror sheet. It is now clear that up to a
simple factor of $\sqrt{\frac{x^+ y^-}{x^- y^+} }$ the dressing factor
in the mirror theory is indeed a pure phase function. More precisely
the combination $\sqrt{\frac{x^- y^+}{x^+ y^-} } \sigma(u,v)$ is a
pure phase in the real axis of the mirror sheet. The same kind of
arguments can be used to prove the reality of the large $L$
Y-functions of \cite{Gromov:2009tv}.

\subsection{Integral representation}
We will show that the dressing phase on the mirror sheet admits some concise integral representation. Based  on that representation we can explicitly see that it has very simple analytical properties. In particular, up to a simple multiplier, namely the simple square root factor identified in the previous section, we can clearly see that this dressing phase is indeed a pure phase function.

\subsubsection{A new representation of the dressing kernel in (mir,mir) kinematics}
We will start form the DHM integral representation \cite{Dorey:2007xn} for $\sigma(\px(u \pm i/2),\px(v \pm i/2))$,
\beq
\sigma\equiv
\exp\[i\chi^{++}+i\chi^{--}-i\chi^{+-}-i\chi^{-+}\]
\eeq
where $\chi^{\pm\pm}=\chi(u\pm i/2,v\pm
i/2)$,
\beqa\label{DHM}
\chi(u,v)
\!\!\!\!&=&\!\!\!\!\frac{1}{i}\oint\limits_{|z_1|=1} \frac{dz_1}{2\pi}\oint\limits_{|z_2|=1} \frac{dz_2}{2\pi}
\frac{1}{z_1-\px(u)}
\frac{1}{z_2-\px(v)}
\log\frac{\Gamma(i w_1-i w_2+1)}{\Gamma(i w_2-i w_1+1)}
\eeqa
and $w_{1,2}=g(z_{1,2}+1/z_{1,2})$. This representation is  valid for the physical kinematics and in particular for $u,v$ in the region $p_1$. Since $p_1=m_1$ we can start with the same expression with $\px(u)$ and $\px(v)$ replaced by $\mx(u)$ and $\mx(v)$ for $u$ and $v$ in the region $m_1$, above the upper cut.

For the kernel ${\cal S}(u,v)\equiv\frac{1}{2\pi i}\d_v\log \sigma(u,v)$ appearing in our TBA equations we have
\beqa\nn
{\cal S}(u,v) &=&-\int_{-2g}^{2g}\int_{-2g}^{2g}\lb
{\cal R}^{(10)}(u,w_1-i0)-{\cal B}^{(10)}(u,w_1+i0)
\rb {\cal G}(w_1-w_2)\times\\
&&\times\lb
{\cal R}^{(01)}(w_2-i0,v)-{\cal B}^{(01)}(w_2+i0,v)\rb dw_1 dw_2\;
\eeqa
where
\beq
{\cal G}(u)\equiv
\frac{\d_{u}}{2\pi i}\log\frac{\Gamma(1-i u)}{\Gamma(1+i u)} = \sum_{a=1}^\infty\left(K_{2a}-\frac{1}{a\pi}\right)+\frac{\gamma
}{\pi}\;.
\eeq
Let us briefly recall how to derive this representation for the dressing kernel from the integral representation (\ref{DHM}). First the pole terms $1/(z_1-x(u))$ and $1/(z_2-x(v))$ are written as derivatives of log's which will give rise to the $\mathcal{B}$'s and $\mathcal{R}$'s in this expression (in this section we often omit the lower indices of $\mathcal{B}$'s and $\mathcal{R}$'s in which case they are equal to \(\dots_{11}\)). The extra derivative to make a kernel out of the phase can also be transported to the log of gamma function by integration by parts and this generates the function $\mathcal{G}$.
The integration contour around the unit circle in the $z_{1,2}$ variable is mapped to an integral from $2g$ to $-2g$ slightly above the real axis and then back from $-2g$ to $2g$ slightly below the real axis for the variable $w_{1,2}$. When $w_1$ is above the real axis we have $z_1=\px(w_1)=\mx(w_1)$ but when we are below we have $z_1=\px(w_1)=1/\mx(w_1)$. This explains why we get that combination of $\mathcal{R}$'s and $\mathcal{B}$'s in the last
formula.

Now that we have transformed the original contour integrals into usual integrals in the real axis we can further replace the integration limits in this expression by $\mp \infty$ because for $|w_1|>2g$ we have ${\cal R}^{(10)}(u,w_1-i0)-{\cal B}^{(10)}(u,w_1+i0) =0$ and similarly for $w_2$. Hence we arrive at the following integral representation for the dressing kernel in the mirror kinematics
\beqa\nn
{\cal S}(u,v) &=&-\int_{-\infty}^\infty\int_{-\infty}^\infty\lb
{\cal R}^{(10)}(u,w_1-i0)-{\cal B}^{(10)}(u,w_1+i0)
\rb {\cal G}(w_1-w_2)\times\\
&&\times\lb
{\cal R}^{(01)}(w_2-i0,v)-{\cal B}^{(01)}(w_2+i0,v)\rb dw_1 dw_2\; \la{repp} \,.
\eeqa
Recall that this expression is derived for $u$ and $v$ in the region $m_1$.
The reason for which we cannot use this integral representation everywhere
on the mirror sheet is the presence of poles of $\cal R$ under the integral
at \(u=\pm w_1\pm i/2-i0\).
To get rid of them we use the relations\footnote{It is often useful to change from $\mathcal{R}$'s to $\mathcal{B}$'s because the latter are much more regular than the former. In particular, since $\im \mx (u),\im \mx (v)>0$ we can never have $1/\mx(u)=\mx(v)$ and thus $\mathcal{B}$ will be pole free when both variables are taken in the mirror sheet. Similarly, in the physical sheet, $|\px (u)|,|\px (v)|>1$ and again $\mathcal{B}$ is regular. Only when $u$ and $v$ are in different kinematics we should worry about regularity of the $\mathcal{B}$ functions.}
 \({\cal R}^{(10)}=K_1-{\cal B}^{(10)}\)
,   \({\cal R}^{(01)}=K_1-{\cal B}^{(01)}\)
and then evaluate the integrals with $K_1$ by poles using\footnote{in these formulae, $w$ is the variable being integrated over in the last convolution}
\beqa
&&\nn
\int K_1(u-w_1)
{\cal G}(w_1-w_2)
{\cal B}^{(01)}\left(w_2+i0,v\right)dw_1 dw_2
=
-{\cal B}^{(01)}\left(u+i/2,v\right)
-\frac{i}{2}{\cal P}^{(1)}(v)\;,\\
&&\nn
\int{\cal B}^{(10)}\left(u,w_1+i0\right){\cal G}(w_1-w_2)K_1(w_2-v)dw_1dw_2
=
-{\cal B}^{(10)}\left(u,v+i/2\right)\;,
\eeqa
where
\beq
{\cal P}^{(a)}(v)=-\frac{1}{2\pi}\d_v\log\frac{\mx(v+ia/2)}{\mx(v-ia/2)}\;.
\eeq In this way, we  get the following representation
valid everywhere in $m_1,m_2,m_3$ for both variables
\beqa
{\cal S}(u,v)
&=&\nn
-{\cal B}^{(11)}(u,v)-\frac{i}{2}{\cal P}^{(1)}(v)+\int_{-\infty}^{\infty}\int_{-\infty}^{\infty}
\left[\lb{\cal B}^{(10)}\left(u,w_1+i0\right)-{\cal B}^{(10)}\left(u,w_1-i0\right)\rb\right.\times
\\
&&\left.\times{\cal G}(w_1-w_2)
\lb{\cal B}^{(01)}\left(w_2+i0,v\right)-{\cal B}^{(01)}\left(w_2-i0,v\right)\rb\right]dw_1dw_2\;.
\eeqa
We see that the integrals can be combined to a contour integral around the
cuts
$(-\infty,-2g)\cup(2g,\infty)$!
This implies that we can write the result in a  fashion similar to \eq{DHM}. Introducing
\beqa
\nn
\hat\chi(u,v)
&\equiv& \frac{1}{i}\int_{|z_1|>1} \frac{dz_1}{2\pi}\int_{|z_2|>1} \frac{dz_2}{2\pi}
\[
\frac{1}{(z_1-\mx(u))}
-\frac{1}{(z_1-\overline\mx(u))}
\]\times\\
&&\times\[
\frac{1}{(z_2-\mx(v))}
-\frac{1}{(z_2-\overline\mx(v))}
\]
\log\frac{\Gamma(i u_1-i u_2+1)}{\Gamma(i u_2-i u_1+1)}
\eeqa
with the integration going along the part of the real axes over $(-\infty,-1)\cup(1,\infty)$.
Then for the physical dressing factor, analytically continued to the mirror in both variables
we get the following representation
\beq
\sigma^{m,m}(u,v)=\frac{1-1/(x^- y^+)}{1-1/(x^+ y^-)}\;\hat\sigma(u,v)\;\;,\;\;
\hat\sigma\equiv
\exp\[i\hat\chi^{++}+i\hat\chi^{--}-i\hat\chi^{+-}-i\hat\chi^{-+}\]\;,
\eeq
where $x=\mx(u),\;y=\mx(v)$ and $\hat\chi^{\pm\pm}=\hat\chi(u\pm i/2,v\pm
i/2)$. We see that the second factor $\hat\sigma$ has the same properties
under the fusion procedure on the mirror sheet as
the physical dressing phase $\sigma$ had on the physical sheet -- one simply
replaces shifts by $\pm i/2$ by $\pm in/2$ for $u$ and  by $\pm im/2$ for $v$ in $\hat \chi$.
Note that $\hat\chi$ is a real function and thus $\hat \sigma$ is a pure phase.
Thus $\hat \sigma(u,v)$ is nothing but the dressing phase of the mirror theory!

Finally let us present yet another interesting representation of the dressing phase in the mirror kinematics.  It is easy to see that ${\cal R}^{(10)}(u,w)$ and ${\cal R}^{(01)}(w,v)$,
as functions of $w$, are regular below the real axis.
Moreover  ${\cal B}^{(10)}(u,w)$ and ${\cal B}^{(01)}(w,v)$ are regular
on the whole complex plane except for the Zhukoswky cuts, see previous footnote.
That implies that the terms with $\cal B B$ and $\cal R R$ in (\ref{repp}) vanish because for those terms we can deform the integration contour to $+i\infty$ and $-i\infty$, correspondingly. For the remaining terms the integration with $\mathcal{G}$ can be done explicitly to yield
\beqa
&&2{\cal S}_{nm}(u,v)-{\cal R}_{nm}^{(11)}(u,v)+{\cal B}_{nm}^{(11)}(u,v)
=
\la{reppp}
-{\cal K}_{n,m}(u-v)-i\mathcal{P}^{(m)}(v)\\
&&-2\sum_{a=1}\int
\left[{\cal B}^{(10)}_{n1}\left(u,w+ia/2\right){\cal B}_{1m}^{(01)}\left(w-ia/2,v\right)
 + {\cal B}^{(10)}_{n1}\left(u,w-ia/2\right){\cal B}_{1m}^{(01)}\left(w+ia/2,v\right)\right]dw \nn
\eeqa
where we wrote the result already after fusion, i.e. for the dressing factor between magnon bound states $n$ and $m$. Quite remarkably this combination of kernels, which is precisely the one appearing in the TBA equations contains no cuts apart from those
at $\im(u)=\pm n/2$ in the $u$ plane and $\im(v)=\pm m/2$  for the $v$ variable, precisely as expected.
This property was also noticed independently in \cite{Arutyunov:2009kf}.

\subsubsection{A new representation of the dressing kernel in the (mir,ph) kinematics}
In this section we analyse the dressing kernel when $\sigma(u,v)$ when the first variable $u$ takes values in the mirror sheet while the second variable $v$ lives in the physical sheet. This is precisely the case for the free terms (without convolutions) in the TBA equations. For example, in
\eq{eqlast} the term $\Phi$ contains $S(u)=\prod_j\s(u,u_{4,j})$ where $u_{4,j}$ are the Bethe roots of the physical theory while $u$ is in the mirror kinematics.
The derivation of a nice integral re presentation for this dressing factor goes along the same lines as in the previous section. We find
\beqa
\log S(u)=-\left[{\cal B}^{(10)}(u,w+i0)-{\cal R}^{(10)}(u,w-i0)\right]*{\cal
G}*\left[\log\frac{B^{(+)}(u+i0)}{B^{(-)}(u+i0)}
-\log\frac{R^{(+)}(u-i0)}{R^{(-)}(u-i0)}\right]
\eeqa
where $R$ and $B$ are defined like in \eq{RBQ} with $x(u)=\mx(u)$ and $x_j^\pm = \px(u_j\pm i/1)$.
As in the previous section this relation is derived in the region $m_1=p_1$ and the next step is to transform this expression in such a way that it allows for a trivial analytical continuation to the full mirror sheet for the $u$ variable. Actually
the r.h.s. is not singular in $m_1,m_2$ (but not in $m_3$),
and thus should coincide with analytical continuation
of the dressing factor. Next we recall that
\beq
B^{(+)}(u_j-i/2)=0\;\;,\;\;R^{(-)}(u_j+i/2)=0
\eeq
while $\log B^{(-)}$ and $\log R^{(+)}$
are regular in $m_1,m_2,m_3$. Assuming $u$ to be real
we can again drop $BB$ and $RR$ terms
and convert $\cal R$ to $\cal B$ as in the previous section to obtain
\beqa
\log S(u)&=&
{\cal B}^{(10)}(u,w+i0)*{\cal G}
*\log\frac{R^{(+)}(u-i0)}{R^{(-)}(u-i0)}
\\&-&{\cal B}^{(10)}(u,w-i0)*{\cal G}*\log\frac{B^{(+)}(u+i0)}{B^{(-)}(u+i0)}
+K_1*{\cal G}*\log\frac{B^{(+)}(u+i0)}{B^{(-)}(u+i0)}\nn
\eeqa
The last term can be computed explicitly, $K_1*{\cal G}*\log\frac{B^{(+)}(u+i0)}{B^{(-)}(u+i0)}
=-\log\frac{B^{(+)}(u+i/2)}{B^{(-)}(u+i/2)}-\sum_j\frac{1}{2}\log \frac{x_j^+}{x_j^-}$,
and in this way we obtain the following integral representation valid in the full mirror sheet $m_1$, $m_2$ and $m_3$
\beqa
\log S&=&
\log\frac{B^{(-)+}}{B^{(+)+}}+\sum_j\frac{1}{2}\log \frac{x_j^+}{x_j^-}\\
\nn&+&
\left({\cal B}^{(10)}(u,w+i0)*{\cal G}*\log\frac{R^{(+)}(u-i0)}{R^{(-)}(u-i0)}
+{\cal B}^{(10)}(u,w-i0)*{\cal G}*\log\frac{B^{(-)}(u+i0)}{B^{(+)}(u+i0)}\right)
\eeqa
Fusion is again trivial and yields
\beqa\la{Smp}
\sum_{k=-\frac{m-1}2}^{\frac{m-1}2}&&\!\!\!\!\!\!\!\!\!\!\!\!\!\!\!\log S(u+ik)=
\sum_{k=-\frac{m-1}{2}}^{\frac{m-1}{2}}\log\frac{B^{(-)}(u+i/2+i
k)}{B^{(+)}(u+i/2+ik)}+\sum_j\frac{1}{2}\log \frac{x_j^{[+m]}}{x_j^{[-m]}}
\\
\nn&+&
\left({\cal B}_{m1}^{(10)}(u,w+i0)*{\cal G}*\log\frac{R^{(+)}(\tilde u-i0)}{R^{(-)}(\tilde u-i0)}
+{\cal B}^{(10)}_{m1}(u,w-i0)*{\cal G}*\log\frac{B^{(-)}(\tilde u+i0)}{B^{(+)}(\tilde u+i0)}\right)\;.
\eeqa
Using the same arguments as in the previous section we could explicitly eliminate $\mathcal{G}$ and one of the convolutions in this representation at the expense of introducing an extra infinite sum over $a$. In this way we could derive an alternative integral representation very similar to that in (\ref{reppp}).

In the next section, the  reality property is discussed in further detail and in particular we explain why the \(Y\)-functions
which solve our integral equations are indeed real.

\subsection{Reality and Analyticity properties of $Y$'s}
We can easily check using the explicit large $L$ solution for the $Y$-functions presented in \cite{Gromov:2009tv} together with the explicit representation of the dressing kernel derived in the previous section that \textit{all} $Y$-functions are real when $u$ is in the real axis\footnote{For the $Y$-functions $Y_{11}$ and $Y_{22}$ associated to the fermionic roots this property is true for $|u|<2g$. }. To understand that this property actually holds for the $Y$-functions even at finite $L$ we should study the reality of  several kernels in the $TBA$ equations and also the reality of the free term (without convolutions). If both are real then the exact finite $L$ solution for \(Y\)-functions
 will be also real. The reality property is of a particular interest for
the future numerical applications of our equations which can be now  done by iterations
starting from the known  large $L$ solution.

The most complicated equation to analyze is the one for the middle node,
eq.(\ref{Yeq2}), which contains the dressing factor in the (fused) kernel and in the free term. We will focus now only on this equation since the reality of all other equations can be checked trivially. Let us explicit in (\ref{Yeq2})
only the ``\textit{dangerous}" terms: \beqa
\log Y_{{\figm}_{n}}&=& 2{\cal S}_{nm}*\log(1+Y_{\figm_m})+
\sum_{k=-\frac{n-1}{2}}^{\frac{n-1}{2}}i\Phi(u+ik)+2{\cal R}_{n2}^{(10)}*\log(1+Y_{\figp_{ 2}})+\dots
\nn
\eeqa
where the $\dots$ stand for the rest of the terms, which are explicitly real. The reason why we also kept the last term as   \textit{dangerous} (i.e. potentially not real) will become clear below.

Inside the kernel $\mathcal{S}_{nm}$ the only non-real contribution comes from the square root of $-\frac{i}{2}\mathcal{P}^{(m)}$ in (\ref{reppp}) and the dangerous terms coming from the fusion of $\Phi$ are those in the first line of (\ref{Smp}) so that we can re-write the \textit{dangerous} terms in the r.h.s of  (\ref{Yeq2}) as\footnote{To simplify the first line in (\ref{Smp}) we use the identity
$$\sum_{k=-\frac{n-1}{2}}^{\frac{n-1}{2}}\log\[\frac{B^{(-)}(u+i/2+i
k)}{B^{(+)}(u+i/2+i k)}\]^2\!\!
\frac{B^{(+)}(u+i/2+ik)R^{(-)}(u-i/2+ik)}{B^{(-)}(u-i/2+ik)R^{(+)}(u+i/2+ik)}=\log\[\frac{B^{(-)}(u+\tfrac{in}{2})}{B^{(-)}(u-\tfrac{in}{2})}\]^2\frac{Q(u-i\tfrac{n+1}2)Q(u-i\tfrac{n-1}2)}{Q(u+i\tfrac{n+1}2)Q(u+i\tfrac{n-1}2)}$$}
\beqa
\la{2lines} \log Y_{{\figm}_{n}}&=& -i\mathcal{P}^{(m)}*\log(1+Y_{\figm_m})+\sum_j\log \frac{x_j^{+}}{x_j^{-}}\\&&+\log\[\frac{B^{(-)}(u+\tfrac{in}{2})}{B^{(-)}(u-\tfrac{in}{2})}\]^2\frac{Q(u-i\tfrac{n+1}2)Q(u-i\tfrac{n-1}2)}{Q(u+i\tfrac{n+1}2)Q(u+i\tfrac{n-1}2)}+2{\cal R}_{n2}^{(10)}*\log(1+Y_{\figp_{ 2}})+\dots
\nn
\eeqa
%
Now we notice that the first line is nothing but the total corrected momentum (compare with the expression (\ref{eq:Energy}) for the corrected energy) which should vanish due to the string theory level matching constraint!\footnote{the
gauge theory analogue  of level matching is the ciclicity of the trace in
the definition of local gauge invariant operators} Thus, the only danger
stems from both terms in the second line: in fact, the are not real (even though the kernel $\mathcal{R}_{n2}^{(10)}$ is real)
but their combination will be shown to be real.

The reason for the second
term to be not real is that the function \(Y_{\figp_{ 2}}\) contains
pole divergencies on the real axis located precisely at the positions of the Bethe roots $u_j$. This can be seen from the free terms (containing no convolutions) in the TBA equation (\ref{YEQp}). We see that iff $n=2$ then we do get singularities in the real axis coming from the zeros of $R^{(-)}(u+i/2)=0$ which are precisely the Bethe roots $u_j$. The zeros of this function induce, via this integral equation, the poles in   \(Y_{\figp_{ 2}}\).\footnote{For $n>2$ we also have poles for the corresponding $Y$-functions but they will lie away from the real axis.} Analyzing all other TBA equations in a similar way we can  easily see that no other free terms give rise to poles in the real axis for any other $Y$-function.

Now let us explain  why the second line in (\ref{2lines}) is explicitly real. The convolution in the presence of these poles should be understood  as $2{\cal R}_{n2}^{(10)}*\log(1+Y_{\figp_{2}}(u-i0))$ which we can re-write as a principal value integral (which will be of course explicitly real) plus half of each residue of the singularities at the Bethe roots, i.e.\footnote{Notice that the residues at these singularities  depend only on the prefactor of the $\log$ since when integrating by parts we get a log derivative which has always unit residue. }
\beqa
2{\cal R}_{n2}^{(10)}*\log(1+Y_{\figp_{2}}(u-i0))&=&
2{\cal R}_{n2}^{(10)}*_{\rm p.v.}\log(1+Y_{\figp_{2}})+
\log\frac{R^{(-)}(u+\tfrac{in}2)B^{(+)}(u+\tfrac{in}2)}{R^{(-)}(u-\tfrac{in}2)B^{(+)}(u-\tfrac{in}2)} \nn
\eeqa
The last term in this expression is not real. However it can be easily seen that it combines with the first term in the second line of (\ref{2lines}) to give a real contribution!

This concludes our check of reality of all the kernels and free terms in all TBA equations. The reality means that the \(Y\)-functions solving these equations will be real, at least on some stretch of the real \(u\)-axis.

\section{Discussion and conclusions}  \la{six}

The integral equations we present are suitable for the numerical study. In the large $L$ limit we can drop all convolutions containing the black nodes $Y_{\figm_n}$ and recover in this way the large $L$ solutions of \cite{Gromov:2009tv} (we also checked this statement numerically). However, compared with the $Y$-system equation in functional form these equations are of easy numerical implementation and the iteration from the large $L$ solution to the finite $L$ case is now accessible. This numerical approach is currently under investigation.

In conclusion, we derived in this paper the system of non-linear integral equations of the TBA type describing, in principle, the spectrum of the states/operators in the full planar AdS/CFT system, including the low lying ones, such as Konishi operator. Not only these equations confirm our Y-system conjectured in \cite{Gromov:2009tv} but they also give a practical way to the  numerical calculation of  the anomalous dimensions as functions of the coupling \(\lambda\). An alternative,  usually numerically quite efficient,  would be the derivation of the Destri-DeVega type equations along the guidelines presented in \cite{Gromov:2008gj} for the \(SU(2)\) principal chiral field.  In any case, a better understanding of the analytical structure
of these equations is needed for the efficient numerics.

A point which we do not completely understand in detail concerns the role of the so called $\mu$-term contributions in the TBA equations. In particular we might need to pick extra contributions in  (\ref{eq:Energy}) coming from further singularities which might arise in the $Y_{n,0}$ functions. In the large $L$ limit such extra terms could probably be identified with the L\"uscher's $\mu$ term contributions. The role of  these extra contributions, if they
are present at all, needs to be further elucidated.


One more unclear point concerns the underlying $PSU(2,2|4)$ symmetry of the problem. In our approach the starting point is  the string theory in the light cone gauge where this symmetry is broken to $SU(2|2)^2$. It would be extremely interesting to understand how the full superconformal symmetry emerges in the TBA equations.

Interesting  questions yet to be considered concern the derivation of a full set of finite size Bethe equations for {\it any} type of excitations of the theory, again along the lines of  \cite{Gromov:2008gj} as well as the generalization of these TBA equations to another integrable example of the AdS/CFT
correspondence, the ABJM duality \cite{ABJM} (see \cite{MZ2,GVcurve,GVall} and references therein for the integrability related works on this theory).

The set of TBA equations derived here should give us access to the full spectrum of AdS/CFT for any coupling.    Hopefully it will help to understand  deep physical reasons of the integrability of \(N=4\) SYM theory.  Knowing the exact results always helps  understanding physics.

{\bf Note Added}

After the work on this project was already finished the paper \cite{Bombardelli:2009ns} appeared
where essentially the similar equations for the vacuum have been derived except the corner, fermionic nodes \(Y_{2,\pm 2}\).
The corresponding equation 5.71 proposed in  \cite{Bombardelli:2009ns} appears to be incorrect.
We derive here the correct equation and
also propose the TBA equations for the excited states.\footnote{
In the preprint arXiv:0902.3930v2 of \cite{Bombardelli:2009ns},
which appeared after our preprint arXiv:0902.4458v1 of the current paper, the
eq. 5.16 (equation 5.71 in arXiv:0902.3930v1), associated with $Y_{2,2}$,
was indeed recognized to be incorrect.}

{\bf Note Added for preprint arXiv:0902.4458v3}

When we were preparing  a paper with the results of the new section \ref{reality},
the second version of the paper arXiv:0904.4575 \cite{Arutyunov:2009kf} appeared where a part of
our new results, concerning
the fusion properties for the mirror dressing factor, was established.
We decided  to update our old paper with these new results.
We also stated more explicitly our preferable chose of the contours and branches
in the integral equations. It is consistent with \cite{Ambjorn:2005wa,Arutyunov:2007tc}
and agrees with that of \cite{Bombardelli:2009ns,Arutyunov:2009ur}.
We also show that for that choice the $Y$-functions for excited states have particulary nice
analytic properties and are real.

{\bf Acknowledgments}

The work of NG was partly supported by the German Science Foundation (DFG) under
the Collaborative Research Center (SFB) 676 and RFFI project grant 06-02-16786
and the grant RFFI 08-02-00287. This research was supported in part by the National Science Foundation under Grand No. NSF PHY05-51164. The work  of VK was partly supported by  the ANR grants INT-AdS/CFT (ANR36ADSCSTZ)  and   GranMA (BLAN-08-1-313695) and the grant RFFI 08-02-00287.  We  thank N.Beisert, M.Staudacher and Z.Tsuboi  for discussions. We acknowledge discussions with G. Arutyunov and S.Frolov about the integrations contours in the integral equations.  VK and PV thank the Kavli Institute for Theoretical Physics of Santa Barbara University, where a part of the work was done, for the kind hospitality.

\end{document}